\documentclass[english,aps,prl,twocolumn,superscriptaddress,amsmath]{revtex4}

\usepackage[T1]{fontenc}
\usepackage[latin9]{inputenc}
\usepackage{color}
\usepackage{units}
\usepackage{amssymb}
\usepackage{graphicx}
\usepackage{esint}
\usepackage{bm}
\usepackage{natbib}
\usepackage{amsmath}
\usepackage{physics}

\usepackage{pdfpages}      %% for adding a pdf file

\usepackage{comment}

\usepackage{ulem}

\setcitestyle{journalcolor= blue}

\usepackage{babel}

\begin{document}

\title{Electric field tunable edge transport in Bernal stacked trilayer graphene}

\author{Saurabh Kumar Srivastav$^{\dagger}$}
\affiliation{Department of Physics, Indian Institute of Science, Bangalore 560012, India}
\author{Adithi Udupa}
\affiliation{Centre for High Energy Physics, Indian Institute of Science, Bangalore 560012, India}
\author{K. Watanabe}
\affiliation{National Institute of Material Science, 1-1 Namiki, Tsukuba 305-0044, Japan}
\author{T. Taniguchi}
\affiliation{National Institute of Material Science, 1-1 Namiki, Tsukuba 305-0044, Japan}
\author{Diptiman Sen}
\affiliation{Centre for High Energy Physics, Indian Institute of Science, Bangalore 560012, India}
\author{Anindya Das$^{\dagger}$}
\affiliation{Department of Physics, Indian Institute of Science, Bangalore 560012, India}

\begin{abstract}
This letter presents a non-local study on the electric field tunable edge transport in an hBN-encapsulated dual-gated Bernal stacked (ABA) trilayer graphene across various displacement fields ($D$) and temperatures ($T$). Our measurements revealed that the non-local resistance ($R_{NL}$) surpassed the expected classical ohmic contribution by a factor of at least two orders of magnitude. Through scaling analysis, we found that the non-local resistance scales linearly with the local resistance ($R_{L}$) only when the $D$ exceeds a critical value of $\sim0.2$ V/nm. Additionally, we observed that the scaling exponent remains constant at unity for temperatures below the bulk-band gap energy threshold ($T<25$ K). Further, the value of $R_{NL}$ decreases in a linear fashion as the channel length ($L$) increases. These experimental findings provide evidence for edge-mediated charge transport in ABA trilayer graphene under the influence of a finite displacement field. Furthermore, our theoretical calculations support these results by demonstrating the emergence of dispersive edge modes within the bulk-band gap energy range when a sufficient displacement field is applied.

\end{abstract}

\maketitle 

The emergence of gapless edge modes at the physical boundaries of a two-dimensional system is one of the most fascinating phenomena in condensed matter physics. Usually, these edge modes are related to the bulk topological order of the system~\cite{thouless1982quantized,hatsugai1993chern,qi2006general,kane2005z,fu2006time,moore2007topological} and play a significant role in electronic transport. Some notable examples are the helical edge modes in $Z_2$ topological insulators~\cite{kane2005quantum,PhysRevLett.96.106802,bernevig2006quantum,onoda2005spin}, chiral quantum Hall edge modes~\cite{laughlin1981quantized,halperin1982quantized}, valley-helical edge modes in graphene~\cite{martin2008topological,qiao2011electronic,zhang2013valley,li2016gate,li2018valley}, kink states\cite{martin2008topological,jung2011valley},
and so on. These edge modes are also believed to be key ingredients for the observation of the electric field-induced magnetism~\cite{son2006half,son2006energy}, valley-dependent transport~\cite{li2016gate,li2018valley}, and half-metallic behavior in graphene or multilayer graphene~\cite{son2006half,son2006energy}.

\begin{figure*}
\centerline{\includegraphics[width=1.0\textwidth]{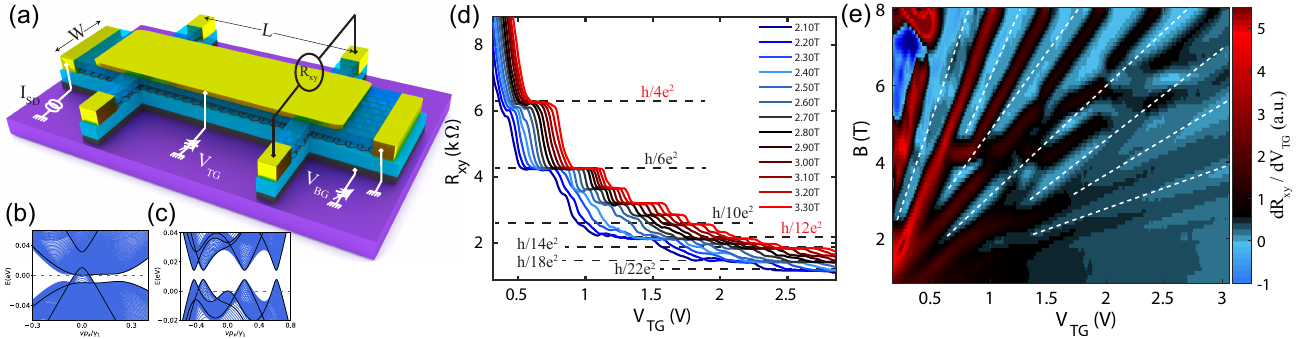}}
\caption{\textbf{(a)} Schematic of the device configuration: Bernal stacked trilayer graphene (TLG) is encapsulated between two hBN substrates and is gated by a graphite back gate ($V_{BG}$) and a metal top gate ($V_{TG}$). \textbf{(b)} and \textbf{(c)}, Band dispersions of Bernal stacked TLG at zero and finite ($\Delta= 200$ meV) displacement field, respectively. \textbf{(d)} $R_{xy}$ response of the device as a function of $V_{TG}$ for several values of the magnetic field. \textbf{(e)} $\frac{dR_{xy}}{dV_{TG}}$ is plotted as a function of $V_{TG}$ and magnetic field. The observation of the crossing points between the different Landau levels, depicted as discontinuities in the QH plateau structure (blue strips) along the white dashed lines, confirms the ABA character of the TLG.}
\label{Figure1}
\end{figure*}

Recently, trilayer graphene (TLG) has emerged as a novel two-dimensional material, 
where several electronic phases, for example, spin-polarized half-metal~\cite{zhou2021half}, spin and valley polarized quarter metal~\cite{zhou2021half}, superconductivity~\cite{zhou2021superconductivity}, correlated Chern insulators and ferromagnetism~\cite{chen2020tunable}, have been realized experimentally. To completely understand the electronic properties of these phases, it is essential to study both bulk and edge transport. Usually, in the absence of a perpendicular displacement field, the band structure of Bernal stacked TLG is described by a set of linear and quadratic bands, which are similar to the low-energy bands of single and bilayer graphene, respectively, as shown in Fig.~\ref{Figure1} (b). However, under the application of the large displacement field, the interplay of layer asymmetry and trigonal warping leads to the formation of new sets of Dirac cones, as shown in Fig.~\ref{Figure1} (c). In addition to modification in the bulk-band structure, the application of displacement field also induces a non-trivial valley Hall state, where the energy gap at the emergent Dirac points is filled by chiral edge modes which propagate in opposite directions between two valleys~\cite{morimoto2013gate}. Although the emergence of the new sets of the Dirac cones has been experimentally observed in a quantum capacitance measurement~\cite{PhysRevLett.121.167601}, an experimental manifestation of the predicted edge modes~\cite{morimoto2013gate} is still lacking.

Non-local transport measurements have been widely used to study the unconventional transport mechanism in two-dimensional systems like the detection of bulk spin and valley transports~\cite{balakrishnan2013colossal,balakrishnan2014giant,PhysRevB.92.161411,PhysRevB.91.165412,gorbachev2014detecting,sui2015gate,PhysRevX.7.031043}. Along with that, the non-local resistance measurement is believed to be an important tool to probe the edge states in the topological insulator~\cite{protogenov2013nonlocal,PhysRevB.84.121302,PhysRevLett.112.026602} and has been widely used to explore the edge transport mechanism in several electronic systems~\cite{roth2009nonlocal,fei2017edge} including the twisted bilayer system~\cite{wang2022bulk}.

In this work, we have carried out non-local resistance measurements in an hBN-encapsulated dual-gated ABA TLG device. The measured non-local resistance was found to be at least two orders of magnitude larger than the classical ohmic contribution. More importantly, the non-local resistance $R_{NL}$ scales linearly with the local resistance $R_L$, suggesting that the charge 
transport is edge-mediated~\cite{PhysRevLett.112.036802,PhysRevB.84.121302,PhysRevLett.112.026602,PhysRevLett.114.096802,PhysRevLett.107.136603,shimazaki2015generation}. The scaling exponent $\alpha$ $(R_{NL}\propto R_{L}^{\alpha})$ was found to be close to $1$ beyond 
a critical displacement field $D$. Below the critical field, the edge states are not dispersive and do not contribute significantly to non-local transport. Similarly, for temperatures ($T$) smaller than the scale of the band gap, $\alpha$ remains close to $1$. On further increasing the temperature, $\alpha$ starts deviating from $1$ due to the contribution of bulk transport. Moreover, we have measured the $R_{NL}$ at different distances between the injected and measured probes, and found that the value of $R_{NL}$ decreases in a linear fashion as the distance $L$ between the probes increases. This is expected for the edge-mediated transport as described in~\cite{shimazaki2015generation}. To further establish our findings, we perform a theoretical calculation of the edge-mode dispersion for different displacement fields, and found that the edge modes become dispersive only above a critical displacement field, which is consistent with our experimental findings.

For the non-local resistance measurement, we fabricated a hBN-encapsulated Bernal stacked TLG device using the standard dry transfer technique with a high mobility of $\sim300000$ cm${^2}$V$^{-1}$s$^{-1}$. Our device is gated by a graphite back gate and a metal top gate. The details of the device fabrication is described in the Supplemental Material\cite{supplement}. Fig.~\ref{Figure1} (a) shows the schematic of the device structure. The electrical resistance was measured using the standard low-frequency lock-in technique. Before discussing details of the non-local measurement, we first discuss the quantum Hall response of the device, which establishes the Bernal stacked trilayer character of the graphene. Fig.~\ref{Figure1} (d) shows a plot of $R_{xy}$ as a function of the top gate voltage $V_{TG}$. The various colour traces correspond to different values of the magnetic field as shown in the legend. One can see that at a low magnetic field, well-developed robust quantum Hall plateaus appear at $h/6e^2$, $h/10e^2$, $h/14e^2$, $\cdots$, which are the characteristic plateaus of TLG~\cite{ezawa2007supersymmetry,koshino2010parity,taychatanapat2011quantum,henriksen2012quantum,yuan2011landau}. We also observe other symmetry broken intermediate plateaus, suggesting the high quality of the device. To further confirm the Bernal stacked trilayer nature of the graphene, in Fig.~\ref{Figure1} (e), we plot a 
two-dimensional color map of $dR_{xy}/dV_{TG}$ as a function of the magnetic field $(B)$ and the top gate voltage $V_{TG}$. The crossing between the Landau levels of the monolayer and bilayer-like bands, whose energies scale differently with the magnetic field, can be seen as discontinuities in the QH plateau structure (blue strips) along the white dashed lines in Fig.~\ref{Figure1} (e). The positions of the crossing points are similar to the earlier experimental observations suggesting the ABA character of the TLG~\cite{taychatanapat2011quantum,campos2016landau,datta2017strong}.

\begin{figure*}
\centerline{\includegraphics[width=1.0\textwidth]{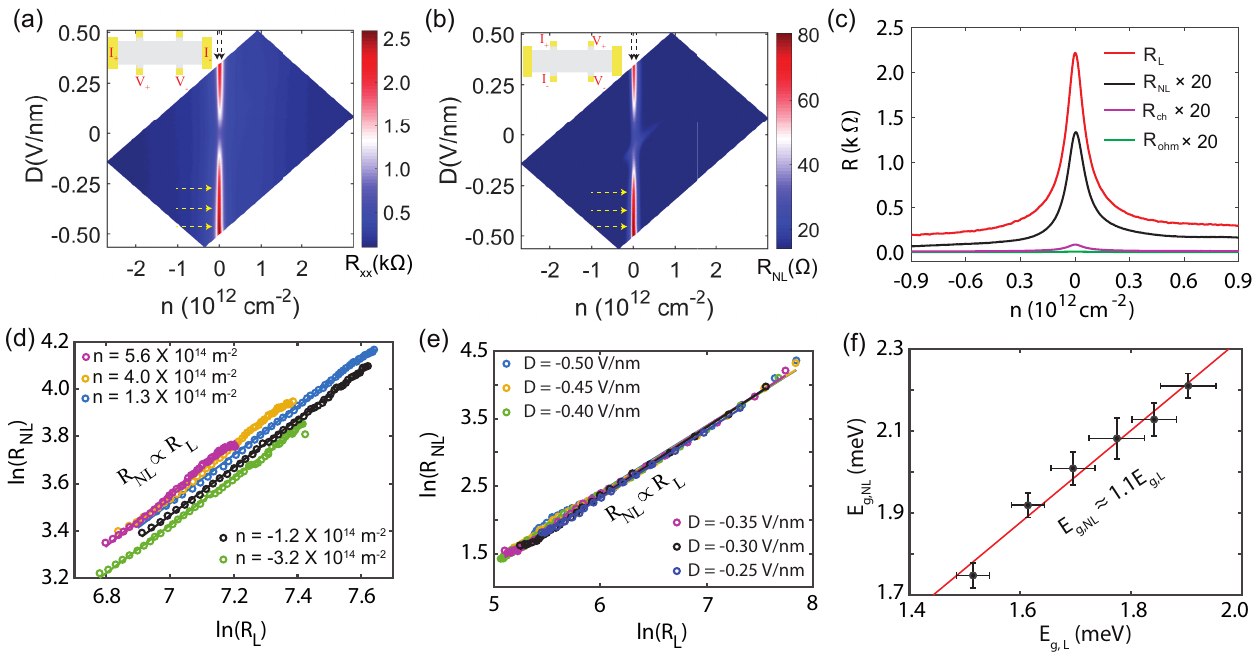}}
\caption{ Color map of $R_L$ \textbf{(a)} and $R_{NL}$ \textbf{(b)} as a function of total carrier density $n$ and the displacement field $D$. (\textbf{c}) The line cuts of $R_L$ (red) and $R_{NL}$ (black) are plotted with density at $D=-0.4$ V/nm. The $R_{NL}$ is multiplied by a factor of 20. The magenta and green curves
(both multiplied by 20) represent the theoretically expected non-local contributions from the charge accumulation (near the edge) and classical ohomic one, respectively. (\textbf{d,e}) Log-log plots of $R_{NL}$ with $R_L$. Open circles are extracted from Figs.~\ref{Figure2} (a,b) for different values of $n$ (along vertical black arrows) near the Dirac point (\textbf{d}) and (\textbf{e}) for different values of $D$ (along horizontal yellow arrows). The solid lines correspond to the linear fitting of the data points with slope $\sim 1$. \textbf{(f)} The activation gap extracted from $R_{NL}$ is plotted versus the gap extracted from $R_{L}$. The red line is the linear fit to these data with slope $\sim 1.1$. Different filled circles correspond to the gap extracted at $D$ ranging from $-0.29$ to $-0.50 $ V/nm. Error bars correspond to the standard deviation associated with the slope of the linear fit.} 
\label{Figure2}
\end{figure*}

The dual gate architecture of the device allows us to tune the carrier density 
$n$ and the displacement field $D$ independently. Fig.~\ref{Figure2} (a) shows a color map of the local resistance $(R_L)$ as a function of the displacement field $D$ and total density $n$. The resistance at the Dirac point at a higher 
displacement fields do not change significantly, suggesting a small band gap of the ABA TLG consistent with the earlier observations~\cite{henriksen2012quantum,datta2018landau,koshino2009gate}. Similarly, Fig.~\ref{Figure2} (b) shows a color map of the non-local resistance $R_{NL}$, (defined as $V_{NL}/I$) as a function of the displacement field $D$ and total density $n$. In Fig.~\ref{Figure2} (c), we plot the line cuts of $R_L$ (red) and $R_{NL}$ (black) as a function of the total density at $D= -0.4$ V/nm. $R_{NL}$ (black) curve is multiplied by 20 to show it on the same resistance scale axis. To rule out the origin of the ohmic contribution due to a classical diffusion of charge transport, we calculate the ohmic contribution using the equation $R_{NL} = \frac{WR_{L}}{\pi L}\text{exp}(-\pi L/W)$,~\cite{gorbachev2014detecting,brune2010evidence,abanin2011giant,balakrishnan2013colossal,sui2015gate} with $L=4$ $\mu$m and $W=1.8$ $\mu$m. The measured $R_{NL}$ is two orders of magnitude larger than the theoretically calculated ohmic contribution, suggesting a non-trivial origin of the observed $R_{NL}$.

Motivated by the earlier non-local resistance measurements in graphene/hBN superlattice and gapped bilayer graphene devices, we perform a scaling analysis of $R_{NL}$ against $R_{L}$. We look for a simple scaling relation $R_{NL}\propto R_{L}^{\alpha}$ to determine the value of $\alpha$. We plot  $\ln R_{NL}$ versus $\ln R_{L}$ in Fig.~\ref{Figure2} (d) as a function of $D$ for different values of $n$ from $-3.2$ to $5.6\times 10^{14}$ m$^{-2}$, and in Fig.~\ref{Figure2} (e) as a function of $n$ for different values of $D$ from $-0.25$ to $-0.50~$V/nm. The data points for these plots are extracted along the vertical dashed black arrows and the horizontal dashed yellow arrows shown in Figs.~\ref{Figure2} (a,b), respectively. The scaling analysis of both Figs.~\ref{Figure2} (d) and (e) show that linear fitting of the plot $\ln R_{NL}$ versus $\ln R_{L}$ gives a slope equal to one ($\alpha \approx 1$).

To further investigate the linear scaling of $R_{NL}$ with $R_L$, we extract the thermal activation gap by measuring the temperature dependence of the local and non-local resistances. As shown in figure S6 in~\cite{supplement}, $R_{NL}$ also follows an activated behavior at high temperatures similar to the $R_{L}$. The temperature dependence of $R_L$ in the activation transport regime is proportional to $e^{E_g/k_{B}T}$. If the non-local resistance follows the scaling relation $R_{NL} \propto R_{L}^{\alpha}$, then its temperature dependence will be proportional to $e^{\alpha E_g/k_{B}T}$. As a result, the activation gap extracted from the non-local resistance $(E_{\text{g,NL}})$ should be $\alpha$ times of the gap obtained from local resistance $(E_{\text{g,L}})$, i.e., $E_{\text{g,NL}} = \alpha E_{\text{g,L}}$. In Fig.~\ref{Figure2} (f), we have plotted the activation gap extracted from non-local resistance against the gap extracted from local resistance. The filled circles correspond to the gaps extracted at displacement fields ranging from -0.29 to -0.50 V/nm. The red line is the linear fit of these data points with slope $\sim 1.1$, again establishing the scaling exponent $\alpha$ close to $1$. Note that though the scaling analysis in Fig.~\ref{Figure2} is limited (the range of $R_L$) due to the small band gap opening in ABA TLG (as seen in Fig.~\ref{Figure2} (f)), further scaling analysis for various displacement fields, temperatures, and channel lengths, which will be discussed in the next section, shows that linear scaling is robust for ABA TLG.

\begin{figure}
\centerline{\includegraphics[width=0.5\textwidth]{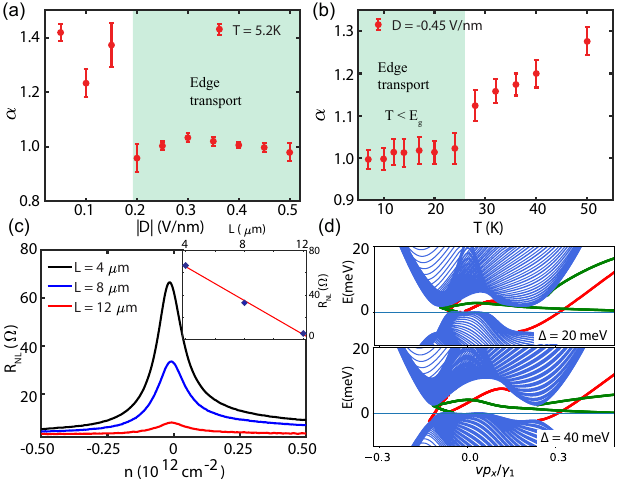}}
\caption{ \textbf{(a)} The scaling exponent $\alpha$ plotted as a function of $D$ at $T=5.2$ K. For $|D|<0.2$ V/nm, $\alpha$ deviates significantly from unity. \textbf{(b)} $\alpha$ plotted as a function of $T$ for $D=-0.45 V/nm$. For $T \gtrsim 25 K$ $\alpha$ starts deviating from unity. \textbf{(c)} $R_{NL}$ is plotted with density for channel lengths of $L=4$ $\mu$m (black), 8 $\mu$m (blue) and 12 $\mu$m (red). Inset: The peak value of $R_{NL}$ (blue circles) plotted for different $L$. The red line is a linear fit. \textbf{(d)} The dispersion of TLG, along with the zigzag edge, is plotted for $\Delta= 20$ meV (top) and $\Delta= 40$ meV (bottom). The green (red) curves correspond to modes on the right (left) edge of the system.} 
\label{Figure3}
\end{figure}

The linear scaling of $R_{NL}$ with $R_L$ resembles the edge-mediated charge transport by the helical edge modes observed earlier in several two-dimensional electronic systems ~\cite{PhysRevLett.112.036802,PhysRevB.84.121302,PhysRevLett.112.026602,PhysRevLett.114.096802,PhysRevLett.107.136603,tiwari2020observation}. Thus, we attribute the linear scaling of $R_{NL}$ with $R_{L}$ to an edge-mediated charge transport in TLG. To strengthen the claim of our findings, we study the effect of the displacement field, temperature and separation between the probes as described below. It can be seen from Fig.~\ref{Figure3} (a) that only above a critical $|D|$ $\gtrsim 0.2$ V/nm the $\alpha$ become close to $1$. Fig.~\ref{Figure3} (b) shows how $\alpha$ varies with $T$ at $D = -0.45$ V/nm, and can be seen that $T \gtrsim 25$ K $\alpha$ starts deviating from $1$. This is consistent with the edge-mediated charge transport in ABA TLG. Above $25$ K, which corresponds to an energy scale similar to the band gap opened in ABA TLG, the bulk states start contributing to the transport, and $\alpha$ deviates from $1$. Fig.~\ref{Figure3} (c) shows $R_{NL}$ for different $L$, and it can be seen from the inset that $R_{NL}$ decreases linearly with $L$, which is in accordance with edge-mediated transport as reported for quantum spin Hall phase and topological insulators~\cite{PhysRevLett.112.036802,PhysRevB.84.121302,PhysRevLett.112.026602,PhysRevLett.114.096802,PhysRevLett.107.136603,shimazaki2015generation}. Note that similar results (critical $D$) are obtained for positive displacement fields.  
However, due to the limited range of $D$ on the positive side, as seen in Fig.~\ref{Figure2} (b), we have presented the data for negative $D$. We have repeated the experiment in different thermal cycles, which is summarized in section-8 in~\cite{supplement}, and shows similar results with scaling exponent 1.

To understand edge-mediated non-local charge transport, we now perform theoretical calculations to confirm the presence of edge states in TLG in the presence of the desired displacement field. We consider a potential drop of $2\Delta$ between the top and bottom layer of TLG due to applying the external displacement field. The magnitude $\Delta$ is related to the experimentally applied displacement field $D$ via the relation $2\Delta=-(\frac{d_{\perp}}{\epsilon_{TLG}}\times D)e$, where $d_{\perp}=0.67$ nm is the separation between the top and bottom layers of TLG, $\epsilon_{TLG}$ is the dielectric constant of the TLG, and $e$ is the electronic charge. Considering the electric field, the Hamiltonian for this system has a form described in detail in~\cite{supplement}. The presence of $\Delta$ opens up a gap in the bulk dispersion but, more interestingly, also gives rise to six new Dirac points each around the Dirac points $K$ and $K'$ points. These play a major role in hosting the edge states in this system.

The Hamiltonian is time-reversal symmetric, implying that the total Hall conductivity $\sigma_{xy}$ summed over all the valleys must be zero. But if we look at a particular valley, the system has a non-zero $\sigma_{xy}^{V}$. We estimate this quantity by numerically calculating the valley Chern number using the method of Fukui et al~\cite{fukui} in the discretized Brillouin zone close to a Dirac point (See \cite{supplement} for details). Although the total Chern number summed over valleys is zero, the valley Chern number $C^V$ equals 2.5 for all the values of $\Delta$ relevant to the corresponding experimental values of $D$. This implies that there is a non-zero valley Hall conductivity of $\sigma_{xy}^V= -2.5 (e^2 /\hbar)$, which agrees with the theoretically predicted value in Ref.~\cite{morimoto2013gate}. The non-zero valley Chern number suggests that there is a possibility of having edge modes in the system. However, the edge modes would not be robust to perturbations since the counter-propagating modes from $K$ and $K^{'}$ valley can hybridize.

In TLG, for $\Delta$ from $20$ meV to $50$ meV, we find that for a zig-zag edge configuration, the edges host gapless modes in the bulk gap. The method used for determining these edge modes is given in \cite{supplement}. Figs.~\ref{Figure3} (d) show plots for $\Delta= 20$ meV and $\Delta= 40$ meV with the edge modes in green (red) for the right (left) edge of the system. Since they are present in the bulk gap, they participate in transport along the edges. However, we note that these edge modes are not protected from backscattering. Hence, there can be intervalley scattering between the states, and a simple dissipative model for edge transport can mimic this and explain the linear scaling between local and non-local resistances as described using a resistor network circuit model in Ref.~\cite{shimazaki2015generation}. Further, the circuit model (equation S20 of Ref.~\cite{shimazaki2015generation}) as explained in section-4 in~\cite{supplement}  captures linear decay of the non-local resistance with the channel length ($L$) as seen in our experiment (the inset of Fig. 3c). We would like to point out that this is unlike the experimental results for bilayer graphene~\cite{shimazaki2015generation}, where the bulk valley Hall effect dominates and gives a cubic relation between $R_{NL}$ and $R_L$ as reported in Refs.~\cite{sui2015gate} and \cite{shimazaki2015generation}. 
From our theoretical calculation, we also find that for small values of the displacement field below $20 ~$meV (See Fig. S10 in \cite{supplement}), the edge modes are approximately flat and non-dispersive, thus not contributing to the non-local charge transport significantly. This is consistent with the experimental results, where we find that as a function of the displacement field, $\alpha$ deviates from one for values of $|D|$ below $0.15$ V/nm ($\Delta <$ 20 ~meV) as shown in Fig.~\ref{Figure3}(b). This is also consistent with Zibrov et al. ~\cite{PhysRevLett.121.167601}, where at the similar displacement field, the Fermi surface undergoes a Lifshitz transition from one electron pocket to multiple isolated Dirac cones.

In general, the non-local signal can originate from mainly three different sources: (i) classical contribution, (ii) a new kind of topological effect - bulk valley Hall effect~\cite{gorbachev2014detecting,sui2015gate,shimazaki2015generation} or (iii) edge transport due to either topological~\cite{morimoto2013gate,brown2018edge,marmolejo2018deciphering} or non-topological (charge accumulation)~\cite{aharon2021long} edge modes. These three mechanisms have also been highlighted in Refs.~\cite{roche2022have,torres2021valley}. Although the non-local measurement is not a smoking gun to distinguish its origin, estimating the non-local contributions from the different sources can help to find its dominant contribution. As shown in Fig.~\ref{Figure2} (c) by the green solid line, the classical ohmic one is ruled out, and similarly, as mentioned before, the linear scaling between $R_{NL}$ with $R_{L}$ in specific parameter spaces of temperature and displacement field, rule out the bulk valley Hall effect. Now, the question is whether the observed edge transport in our experiment originated from a topological or non-topological effect. To figure it out, we estimate the contribution from the non-topological charge accumulation effect ($R_{ch}$)\cite{aharon2021long} and shown by the solid magenta line in Fig.~\ref{Figure2} (c) (detail in section-7 in~\cite{supplement}), which is one order of magnitude smaller than the measured non-local signal (solid black line in Fig.~\ref{Figure2} (c)). Further, the linear decay of $R_{NL}$ with $L$ (Fig.~\ref{Figure3} (c)) rules out the charge accumulation contribution, which would have scaled exponentially with the length\cite{aharon2021long} (see section-7 in~\cite{supplement}). Thus, the dominant contribution to our non-local signal presumably comes from the dispersive edge modes of TLG as predicted in Ref.~\cite{morimoto2013gate} and shown by our theoretical calculation (beyond a critical displacement field). Our findings are in sharp contrast to Ref\cite{aharon2021long} on a non-aligned single-layer graphene device, where the dominant contribution to the non-local signal was the charge accumulation effect~\cite{aharon2021long}, and is expected due to the absence of dispersive edge modes~\cite{PhysRevB.54.17954}.

In conclusion, the consistent linear scaling of non-local resistance across temperature variations, displacement field changes, and a threefold variation in channel length corresponds to the dispersive edge mode transport in correlation with our theoretical calculations.

S.K.S. thanks, Adhip Agarwala, Priya Tiwari, and Manbendra Kuiri, for the useful discussions. S.K.S. acknowledges PMRF, Ministry of Education, India for financial support. D.S. thanks SERB, India for support through project JBR/2020/000043. A.D. thanks the Department of Science and Technology (DST) and Science and Engineering Research Board (SERB), India for financial support (SP/SERB-22-0387) and acknowledges the Swarnajayanti Fellowship of the DST/SJF/PSA-03/2018-19. A.D. also thanks CEFIPRA project SP/IFCP-22-0005.

$^{\dagger}$Corresponding author:\newline
ssaurabh@iisc.ac.in \newline
anindya@iisc.ac.in

% \bibliography{ReferencesMS}{}

%

\onecolumngrid
\newpage
\thispagestyle{empty}
\mbox{}


\section*{Supplementary material}

**SM section-1: Device fabrication and basic characterization.**

We fabricated a dual gated hBN encapsulated stacked trilayer graphene (TLG) device for the non-local resistance measurement using the standard dry transfer technique[1]. The process involved the mechanical exfoliation of graphite and bulk hBN crystal on the SiO$_2$/Si wafer to obtain TLG and thin hBN. TLG was first identified under an optical microscope and was later confirmed via Raman spectroscopy. As shown in Supplemental Fig. S1 (a), we observe two characteristic Raman peaks, which appear around $\sim 1580$ cm$^{-1}$ ('G') and $\sim 2960$ cm$^{-1}$ ('2D') and belong to the "graphene" family. Furthermore. the peak intensity ratio and the spectral decomposition of the '2D' peak into six Lorentzian (Supplemental Fig. S1 (b)) confirms the (ABA) trilayer nature of the graphene flake[2–4]. As discussed in the main manuscript, the ABA character of the used TLG flake was further confirmed in quantum Hall measurements.

The fabrication of the hBN/TLG/hBN/graphite involves the following process. First, a top hBN was picked up at a temperature of $90°C$ using a poly-bisphenol-A-carbonate-coated polydimethylsiloxane block mounted on a glass slide attached to the tip of a micro-manipulator. The picked-up hBN flake was aligned over the TLG flake, which was picked up at a temperature of $90°C$. Following a similar procedure, we picked up the bottom hBN and graphite. The resulting heterostructure (hBN/TLG/hBN/graphite) was dropped down on top of an oxidized silicon wafer ($p^{++}$ doped silicon with SiO$_2$) at a temperature of $180°C$. This final stack was cleaned in chloroform (CHCl$_3$) followed by acetone and isopropyl alcohol (IPA). The next step involved electron-beam lithography (EBL) to define the contact region. Poly-methyl-methacrylate (PMMA) was coated on the resulting heterostructure. The contact regions were defined using EBL. The edge contacts were achieved by reactive ion etching (a mixture of CHF$_3$ and O$_2$ gas was used with a flow rate of 40 and 4 sccm, respectively, at $25°C$ with RF power of 60 W), where the etching time was optimized such that the bottom hBN is not etched completely. This is done to isolate the metallic contacts from the bottom graphite that will serve as the back gate. Next, thermal deposition of Cr/Pd/Au (5/15/60 nm) was performed to make the contacts in an evaporator chamber having a base pressure of $\sim 1 \times 10^{-7}$ to $2 \times 10^{-7}$ mbar and followed by lift-off procedure in acetone and IPA. We again coat the PMMA on the resulting device and define the top gate area using EBL. Finally, we did a thermal deposition of Cr/Au(5/60 nm) followed by the lift-off procedure. The optical image of the final device is shown in Supplemental Fig. S2 (a).

After fabricating the device, we first study the zero-field back gate response of the device, as shown in Supplemental Fig. S2 (b). The measured resistance data is fitted with the equation[5–9]

$$R = R_C + \frac{L}{We\mu\sqrt{n_0^2 + \left(\frac{C_{BG}(V_{BG}-V_{DP})}{e}\right)^2}}, \quad (S1)$$

where $R_C$, $L$, $W$, $\mu$, and $e$ are, respectively, the contact resistance, length, width, mobility, and electron charge. The carrier concentration of the channel is given by $\frac{C_{BG}(V_{BG}-V_{DP})}{e}$ with $C_{BG}$ and $V_{DP}$ being the capacitance per unit area of the bottom graphite gate, and the voltage at the charge neutrality point,

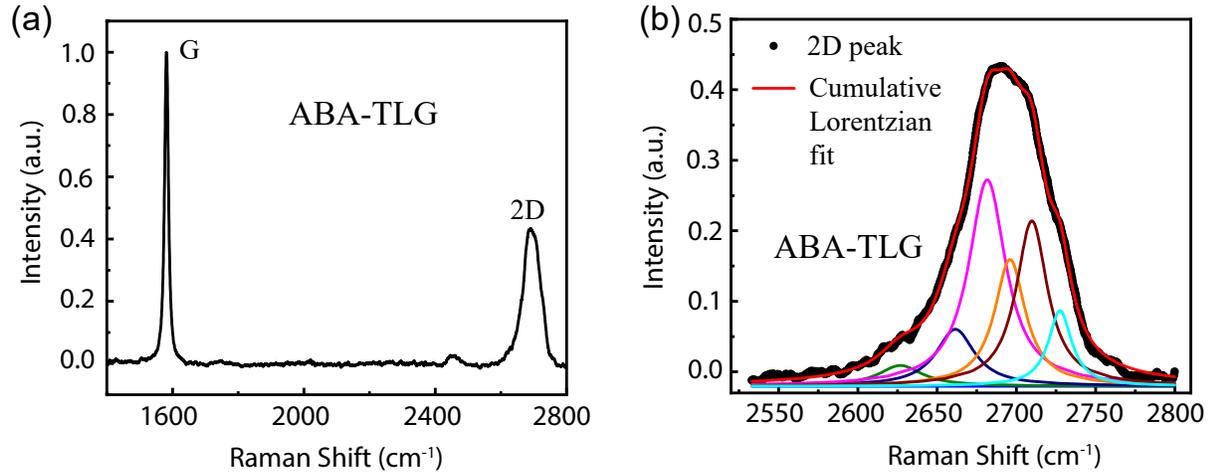

**Fig. S 1: Raman Spectroscopy of Bernal stacked (ABA) trilayer graphene:** (a) Raman spectrum of the used "graphene" flake. (b) Zoomed Raman spectra around 2D peak. It is fitted well with the six characteristic Lorentzian peaks of Bernal stacked (ABA) trilayer graphene.

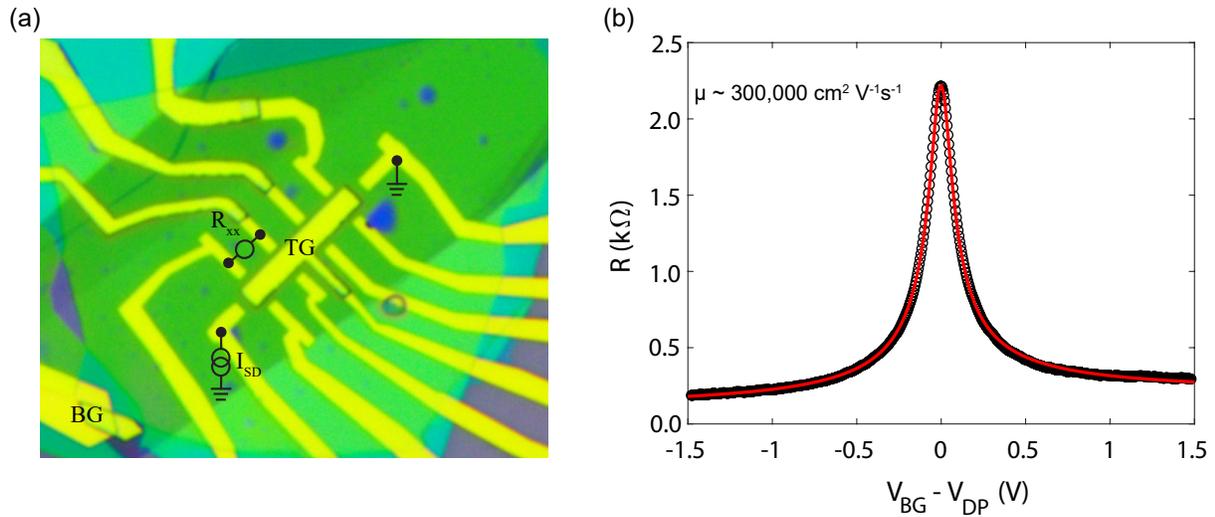

**Fig. S 2: optical image and basic device characterization at zero magnetic fields:** (a) Optical image of the dual gated hBN encapsulated Bernal stacked trilayer graphene. (b) The four-probe resistance (measurement configuration is shown in (a)) is plotted as a function of back gate voltage at $T = 5.2$ K. Open circles show the experimental data and the red curve is the fit of data in accordance with Eq. (S1). From this fit, extracted mobility was found to be $\sim 300{,}000$ cm$^2$V$^{-1}$s$^{-1}$.

respectively, and $n_0$ is the charge inhomogeneity. The extracted mobility from the fitting was found to be $\sim 300{,}000$ cm$^2$V$^{-1}$s$^{-1}$, indicating the high quality of the device.

**SM section-2: Measurement scheme of non-local resistance detection and absence of spurious non-local voltage**

Detecting the true non-local resistance requires a careful design of the measurement setup. One of the most common artifacts that can give rise to spurious non-local resistance is the current leakage to the voltage terminals due to the finite input impedance between the input and ground terminals. The origin and the magnitude of this spurious non-local resistance can be simply understood with the basic schematic shown in Fig. S3 (a). The magnitude of the spurious non-local voltage for the given schematic will be

$$V_{NL,amp}^s = \frac{V_C R_{amp}}{R_{amp} + R_A} - \frac{V_C R_{amp}}{R_{amp} + R_B} \approx (R_B - R_A)\frac{V_C}{R_{amp}}, \tag{S2}$$

where $V_C$ is the potential of the point $C$, and $R_A$ ($R_B$) is the net resistance from point C to the input of voltage metal lead terminal $A$ ($B$). So any difference between $R_A$ and $R_B$ will, in principle, contribute to this spurious non-local voltage signal. Since the typical values of $R_A$ and $R_B$, should be of the order of magnitude of the sample resistance $R_{sample}$, the magnitude of this spurious non-local resistance will be

$$R_{NL,amp}^s \approx (R_B - R_A)\frac{R_{sample}}{R_{amp}}. \tag{S3}$$

It is evident from Eq. S3 that using the high input impedance voltage amplifier $R_{amp}$ will significantly reduce the magnitude of this spurious non-local resistance signal. In our experiment, the non-local voltage was measured with a Stanford Research Systems SR830 lock-in amplifier after being amplified with a commercial differential voltage amplifier SR560 ($R_{amp} = 100 \ M\Omega$) (see Fig. S3 (b)). Since the typical sample resistance in our device was found to be $\sim 2.5$ k$\Omega$, the maximum magnitude of this spurious non-local resistance will be of the order of $0.1\Omega$ or less, which is at least two orders of magnitudes smaller than the observed non-local resistance in our device.

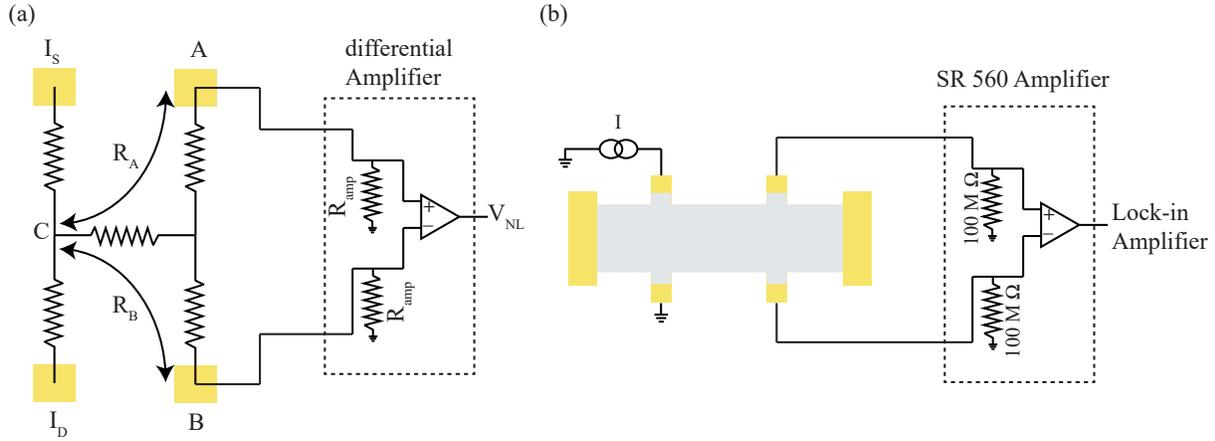

**Fig. S 3: Non-local measurement schematic to eliminate spurious signal from the current leakage:** **(a)** A simple circuit model of the non-local resistance measurement set-up. This simple model is used to estimate the upper bound on the spurious non-local signal due to the current leakage arising from the finite input impedance of the voltage pre-amplifier. $R_A$ ($R_B$) is the total resistance between the point $C$ and the metal lead $A$ ($B$). **(b)** Schematic of the non-local voltage measurement with SR 560 voltage pre-amplifier.

**SM section-3: Absence of heating effect for observed non-local resistance**

To rule out the heating effect [10] as the origin of the observed non-local signal, we measure the current dependence of the non-local resistance. Fig. S4 shows the plot of non-local resistance with density at 20, 50, and 100 nA of the excitation current. The measured non-local resistance was the same for three different excitation currents, confirming the absence of a heating effect as the origin of the observed signal.

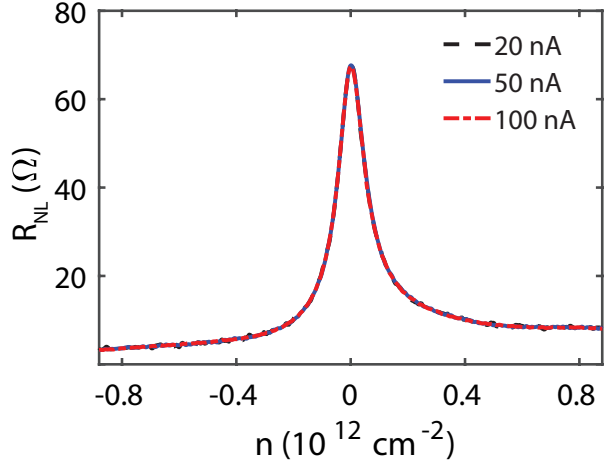

**Fig. S 4: Absence of heating effect for observed non-local resistance:** The non-local resistance is plotted for three different excitation currents of 20, 50, and 100 nA. As the plot shows, the measured signal is independent of the excitation current, ruling out the heating effect as the origin of the observed non-local resistance.

**SM section-4: Length dependence of the non-local resistance**

In Fig. S5 (b), we plot the gate response of the non-local resistance for three different channel lengths ($L$: distance between the injection and detection terminals) of 4 (black), 8 (blue) and 12 $\mu$m (red) as can be seen from the optical image of the device (Fig. S2a). A schematic for the measurement configuration is shown in Fig. S5 (a). As the channel length is increased, the non-local resistance reduces significantly. To understand the functional behaviour of this decrease with the sample length, we plot the peak values of the non-local resistance as a function of the length, as shown in Fig. S5(c) and the data points are best captured by a linear fitting. The linear decrease of the measured non-local signal with the length is in accordance with edge-mediated transport. The length dependence of the non-local resistance has also been observed in the quantum spin Hall phase and topological insulator samples, where the charge transport was dominated mainly by the helical edge modes at the boundary of the sample[11–16]. As discussed later in the theoretical section, there are edge modes in ABA TLG. The edge states are counter-propagating originating from the $K$ and $K'$ valleys. Since the system is non-topological and only has a non-zero valley Hall conductance, the counter-propagating modes are not protected by backscattering into each other. Thus, this transport can be modeled by a dissipative transport along the edges. Such a circuit model has been studied in Ref. 16, and the non-local resistance based on the resistor network is given by[16]

$$R_{NL} \propto \frac{4l_1 l_2}{W(L + l_1 + l_2)} R_L, \tag{S4}$$

where $l_1$ is the distance to the current probe from the nearest end of the sample, $l_2$ is the distance to the voltage probe from the other end, $L$ is the distance between the current and voltage probes, and $W$ is the

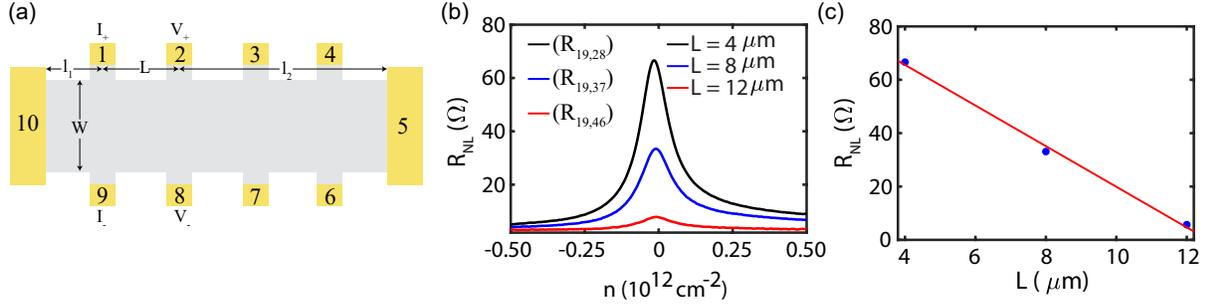

**Fig. S 5: Length dependence of non-local resistance:** (a) Schematic of the non-local resistance measurement for three different channel lengths. The current is injected between contact 1 and 9, and the non-local voltage is measured between contacts 2 and 8 (3 and 7 / 4 and 6) for $4\mu m$ ($8\mu m$/$12\mu m$) of channel length. $l_1$ is the distance to the current probe from the nearest end of the sample, $l_2$ is the distance to the voltage probe from the other end, $L$ is the distance between the current and voltage probes, and $W$ is the width of the device. (b) The non-local resistance is plotted for three different channel lengths of $L = 4$ $\mu m$ m (black, $R_{19,28}$), 8 $\mu m$ (blue, $R_{19,37}$), and 12 $\mu m$ (red, $R_{19,46}$). With increasing channel length $L$, the non-local resistance reduces significantly. The first two subscripts of $R_{19,28}$ correspond to the current injection contacts, while the last two correspond to the voltage probes. (c) The peak value of the non-local resistances shown in (b) is plotted as a function of the length $L$. The red line is the linear fit to these data points. The linear decrease in the non-local resistance with sample length is consistent with edge-mediated non-local charge transport.

width of the device. This simple resistor model captures the $\alpha = 1$ scaling relations between the local and non-local resistances. In our measurement configuration, as shown in Fig. S5 (a), the current injection contacts are always fixed; hence $l_1$ is constant. In the denominator of the Eqn. S4, the term $W(L + l_1 + l_2)$ is constant because as we change the voltage probes, $L$ increases but $l_2$ decreases by the same amount. In the numerator, we can replace $l_2$ by $L' - L$, where $L'$ is the distance from the current injection contact to the other end of the sample, which is fixed in our case. Hence, $R_{NL}$ in Eqn. S4 will be proportional to the length $L$ ($R_{NL} \propto (L' - L)$). This simple model clearly explains the linear length dependence of our non-local resistance. However, we would like to point out that the scaling relation $\alpha = 1$ seen in our experiment is quite different than the experimental results obtained by measurements done on bilayer graphene. In the case of bilayer graphene, despite the presence of zig-zag edge modes, the bulk conduction dominates, giving a cubic relation between $R_{NL}$ and $R_L$ as reported in Refs. 16 and 17.

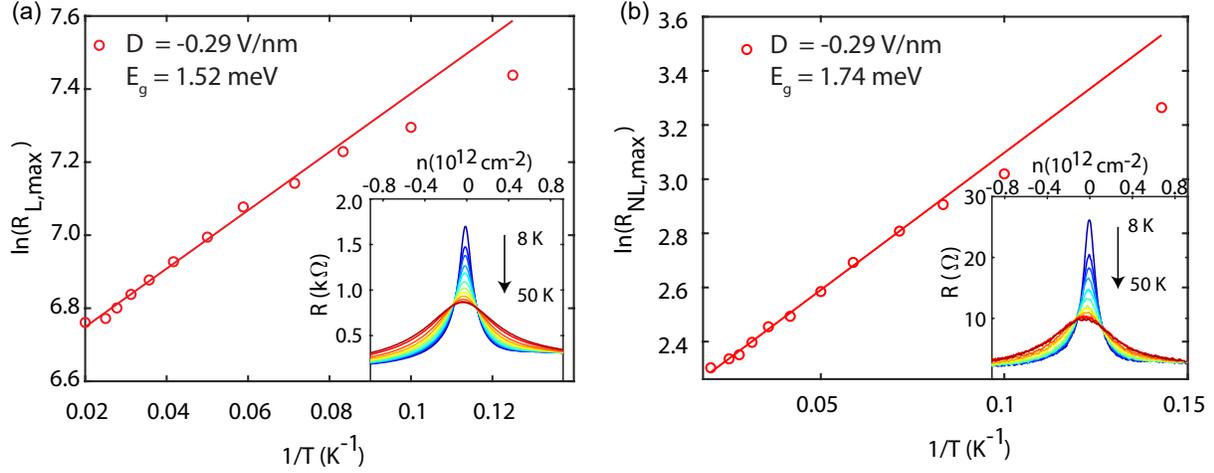

**Fig. S 6: Temperature dependence of local and non-local resistances: (a/b)** The peak resistance values of the local/non-local resistance ($\ln R_{L,\text{max}}$, $\ln R_{NL,\text{max}}$) are plotted as a function of $1/T$ at a displacement field value of $D = -0.29$ V/nm. Open circles represent the experimental data points, while the solid line is the linear fit to these data points to extract the thermal activation gap. The activation gap extracted from the local (non-local) resistance was 1.52 (1.74) meV. Insets show plots of local ($R_L$) and non-local resistance ($R_{NL}$) as a function of the carrier density at $D = -0.29$ V/nm. Different colour traces in **(a,b)** correspond to different values of the temperature ranging from 8 to 50 K.

**SM section-5: Thermal activation gap from the temperature dependence of local and non-local resistances**

To extract the activation gap from the temperature dependence of the local and non-local resistances, we plot the $\ln R_{L,\text{max}}$ and $\ln R_{L,\text{max}}$ (peak resistance value at Dirac point) versus $1/T$. In Fig. S6 (a), $\ln R_{L,\text{max}}$ is plotted versus $1/T$ at $D = -0.29$ V/nm. The open circles represent the experimental data points, and the solid line is the linear fit to the data points to extract the activation gap. The activation gap extracted from the local resistance was found to be 1.52 meV. Similarly, to determine the activation gap from the non-local resistance, $\ln R_{NL,\text{max}}$ is plotted versus $1/T$ at $D = -0.29$ V/nm in Fig. S6 (b). Here also, the open circles represent the experimental data points, and the solid line is the linear fit to the data points to extract the activation gap. The activation gap from the non-local resistance measurement was found to be 1.74 meV. The insets of Figs. S6 (a) and (b) show plots of $R_L$ and $R_{NL}$ as a function of the carrier density at $D = -0.29$ V/nm, respectively. Different colour traces correspond to different temperature values ranging from 8 to 50 K. To extract the displacement field dependence of scaling exponent $\alpha$, we perform a similar measurement at several values of the displacement field.

**SM section-6: Bulk and Edge mode dispersion of Bernal stacked (ABA) trilayer graphene at finite $\Delta$**

The non-local and local resistance have a scaling relation given by $R_{NL} \sim R_L^\alpha$. The measurements show that $\alpha = 1$ for the range of displacement field from 0.2 V/nm to 0.5 V/nm. The value of $\alpha$ stays close to 1 up to a temperature of around $25K$ at which the temperature is close to the band gap. These observations strongly suggest that the transport is dominated in this sample by edge modes. In this section, we show theoretically that for trilayer graphene with displacement fields consistent with the experiments, the valley Chern number is non-zero with a large value of 2.5 for a given valley and spin. This suggests the presence of edge modes in the system, although they are not topologically protected. We show that the system does host zig-zag edge modes for the displacement fields of our interest. A simple resistor circuit model then explains the linear relation between the non-local and local resistance measurements.

**Bulk properties of trilayer graphene in the presence of a perpendicular electric field:** Considering ABA stacked trilayer graphene with $|A_i\rangle$ and $|B_i\rangle$ representing the Bloch functions on the $i^{th}$ layer, the Hamiltonian in the basis $|A_1\rangle, |B_1\rangle, |A_2\rangle, |B_2\rangle, |A_3\rangle$ and $|B_3\rangle$ has the following form close to the Dirac point[18],

$$\mathcal{H}(\boldsymbol{p}) = \begin{pmatrix} H_0 & V & W \\ V^\dagger & H_0' & V^\dagger \\ W & V & H_0 \end{pmatrix},$$

where $H_0 = \begin{pmatrix} 0 & v_0 \boldsymbol{p}^* \\ v_0 \boldsymbol{p} & \delta \end{pmatrix}$, $H_0' = \begin{pmatrix} \Delta' & v_0 \boldsymbol{p}^* \\ v_0 \boldsymbol{p} & 0 \end{pmatrix}$, $V = \begin{pmatrix} -v_4 \boldsymbol{p}^* & v_3 \boldsymbol{p} \\ \gamma_1 & -v_4 \boldsymbol{p}^* \end{pmatrix}$, and $W = \begin{pmatrix} \gamma_2/2 & 0 \\ 0 & \gamma_5/2 \end{pmatrix}$, with $\boldsymbol{p} = p_x + i p_y$ being the momentum as measured from the Dirac point. The different band velocities are given by $v_i = 3a\gamma_i/2\hbar$, where the $\gamma_i$'s are the couplings as described below, and $a \sim 2.46$ Å is the lattice constant, i.e., the distance between the two nearest $A$ atoms in the same layer. The coupling $\gamma_0$ comes from hopping between nearest-neighbour sites within a layer. The parameters $\gamma_1$, $\gamma_3$ and $\gamma_4$ are the nearest interlayer couplings in this model. $\gamma_1$ comes from coupling between the dimer atoms ($B_1 \leftrightarrow A_2$, $A_2 \leftrightarrow B_3$). Since this is a vertical coupling (see Fig. S7), the corresponding matrix element in the Hamiltonian is independent of the momentum $\boldsymbol{p}$. The parameter $\gamma_3$ describes the couplings between the non-dimer orbitals ($A_1 \leftrightarrow B_2$, $B_2 \leftrightarrow A_3$) and $\gamma_4$ between dimer and non-dimer atoms ($A_1 \leftrightarrow A_2$, $B_1 \leftrightarrow B_2$, $A_2 \leftrightarrow A_3$ and $B_2 \leftrightarrow B_3$). These couplings have an in-plane component (Fig. S7)) analogous to the intra-layer nearest-neighbour hopping and therefore come with a momentum-dependent factor in the Hamiltonian. The couplings between the first and third layers have strengths $\gamma_2$ between the $A$ sublattices and $\gamma_5$ between the $B$ sublattices. The parameter $\delta$ is the on-site energy asymmetry coming from the sites which are involved in the $\gamma_1$ couplings and the sites which are not.

9

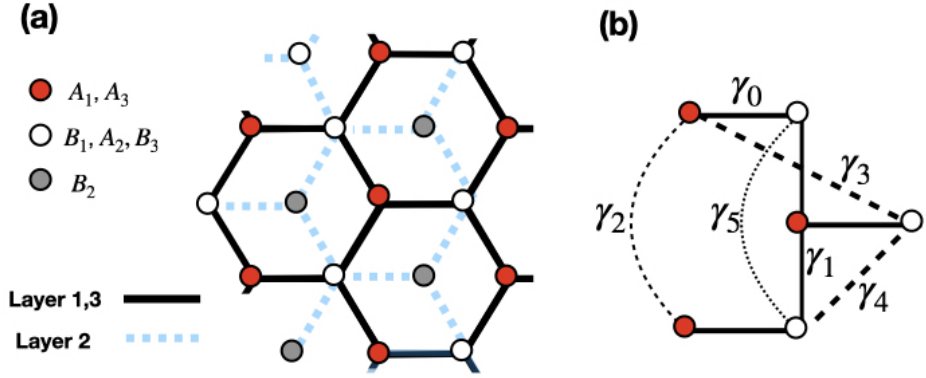

**Fig. S 7:** (a) Lattice structure of ABA stacked trilayer graphene. The first and the third layer sit exactly on top of each other, whereas the second layer has the atoms in the $A$ sublattice right below the atoms in the $B$ sublattice on the first layer. Since these are positioned vertically above each other, the orbitals $B_1$- $A_2$ and similarly $A_2$- $B_3$ are called dimer atoms. (b) Side view of the lattice with the intra- and interlayer couplings marked.

Additionally, in the presence of the displacement field, the Hamiltonian becomes

$$\mathcal{H}(\boldsymbol{p}) = \begin{pmatrix} H_0 + U & V & W \\ V^\dagger & H_0' & V^\dagger \\ W & V & H_0 - U \end{pmatrix}, \quad \text{where} \quad U = \Delta \begin{pmatrix} 1 & 0 \\ 0 & 1 \end{pmatrix}. \tag{S5}$$

We can block diagonalize the Hamiltonian by re-defining the basis as

$$\frac{1}{\sqrt{2}}(|A_1\rangle - |A_3\rangle), \ |B_2\rangle, \ \frac{1}{\sqrt{2}}(|B_1\rangle + |B_3\rangle), \ \frac{1}{\sqrt{2}}(|B_1\rangle - |B_3\rangle), \ |A_2\rangle, \ \frac{1}{\sqrt{2}}(|A_1\rangle + |A_3\rangle). \tag{S6}$$

In this basis, the TLG Hamiltonian takes the form[18],

$$\mathcal{H} = \begin{pmatrix} H_\circ & D_- \\ D_+ & H_\bullet \end{pmatrix}, \tag{S7}$$

where $H_\circ$ ($H_\bullet$) is written in the first (second) three bases of Eq. (S6),

$$H_\circ = \begin{pmatrix} -\gamma_2/2 & 0 & 0 \\ 0 & 0 & -\sqrt{2}v_4\bm{p} \\ 0 & -\sqrt{2}v_4\bm{p}^* & \gamma_5/2 + \Delta' \end{pmatrix},\tag{S8}$$

$$H_\bullet = \begin{pmatrix} -\gamma_5/2 + \delta & 0 & 0 \\ 0 & \delta & -\sqrt{2}v_4\bm{p} \\ 0 & -\sqrt{2}v_4\bm{p}^* & \gamma_2/2 \end{pmatrix},\tag{S9}$$

$$D_+ = \begin{pmatrix} v_0\bm{p} & 0 & \Delta \\ 0 & v_0\bm{p}^* & \sqrt{2}\gamma_1 \\ \Delta & \sqrt{2}v_3\bm{p} & v_0\bm{p}* \end{pmatrix}, \quad \text{and} \quad D_- = D_+^\dagger.\tag{S10}$$

In $D_\pm$, $2\Delta$ is the potential drop between the top and bottom layers of TLG due to the application of the external displacement field via local gating. It is related to the experimentally applied displacement field $D$ via the relation $2\Delta = -(\frac{d_\perp}{\epsilon_{TLG}} \times D)e$, where $d_\perp = 0.67$ nm is the distance between the top and bottom layers of TLG, $\epsilon_{TLG}$ is the dielectric constant of the material, and $e$ is the electronic charge.

We now plot the dispersion obtained for this model using the values of the parameters given in the literature as follows: $\gamma_0 = 3.16$ eV, $\gamma_1 = 0.39$ eV, $\gamma_2 = -0.02$ eV, $\gamma_3 = 0.32$ eV, $\gamma_4 = 0.044$ eV, $\gamma_5 = 0.038$ eV, $\delta = 0.05$ eV, and $\Delta = 0.2$ eV. From Fig. S9, we can clearly see that the bulk dispersion of TLG contains a single Dirac cone and six other off-centred Dirac cones, which are emerging due to the interplay between the trigonal warping and the layer asymmetry term $\Delta$. The emergent new Dirac cones in the bulk band dispersion were observed experimentally in quantum capacitance measurements of a high-quality, Bernal stacked TLG sample[19].

**Valley Chern number calculation for trilayer graphene** The Hamiltonian in Eq. (S7) is time-reversal symmetric. Thus the total Hall conductivity $\sigma_{xy}$ summed over the different valleys must be zero. But if we look at a particular valley, we find that this system has a non-zero $\sigma_{xy}^V$. We estimate this quantity using the method of Fukui et al.[20]. For a system with $n$ bands and $|\psi(\bm{k})\rangle$ being the normalized wave functions in the $n^{th}$ band, the total Chern number is given by

$$C = \sum_{n,occ} \frac{1}{2\pi} \int_{BZ} d\bm{k}\, (F_n)_z,\tag{S11}$$

11

where the summation runs over all the occupied bands. The quantity $(F_n)_z$ is called the Berry curvature and is equal to the curl of the Berry connection $\boldsymbol{A}^n = (\boldsymbol{A}_x^n, \boldsymbol{A}_y^n)$. These quantities are given by

$$\boldsymbol{A}^n = -i\Big( \langle \psi_n(\boldsymbol{k})| \partial_{k_x} |\psi_n(\boldsymbol{k})\rangle, \langle \psi_n(\boldsymbol{k})| \partial_{k_y} |\psi_n(\boldsymbol{k})\rangle \Big), \tag{S12}$$

and therefore

$$(F_n)_z = (\nabla \times \boldsymbol{A})_z = -i\Big(\frac{\partial}{\partial k_x}A_y^n - \frac{\partial}{\partial k_y}A_x^n\Big). \tag{S13}$$

We are interested in evaluating the valley Chern number numerically. Instead of integrating over the entire Brillouin zone, we take a discretized segment of the $(k_x, k_y)$ space around one of the Dirac points ($K$ or $K'$) about which we would like to calculate $\sigma_{xy}^V$. We ensure that this space contains the Dirac point and the six-off centred Dirac points of the trilayer graphene. Since most of the contribution to the Berry curvature (Fig. S8 (b)) comes from these points, this is a valid region to work with. We then divide this region into smaller plaquettes (see Fig. S8 (a)) and use Fukui's method to calculate $\sigma_{xy}^V$ as follows. For the $n^{th}$ band, we define a quantity

$$U_{d\boldsymbol{k}}^n(\boldsymbol{k}) = \langle \psi_n(\boldsymbol{k}) | \psi_n(\boldsymbol{k}+d\boldsymbol{k}) \rangle, \tag{S14}$$

where $d\boldsymbol{k}$ takes us to the next point in the chosen plaquette. The quantity $F_n(\boldsymbol{k})$ is now written in terms of $U_n(\boldsymbol{k})$ as

$$F_n(\boldsymbol{k}) = -i \, \log\Big(U_{dk_x}^n(\boldsymbol{k}) U_{dk_y}^n(\boldsymbol{k}+dk_x) U_{dk_x}^{n^{-1}}(\boldsymbol{k}+dk_y) U_{dk_y}^{n^{-1}}(\boldsymbol{k})\Big). \tag{S15}$$

To understand this quantity better, consider the grid shown in Fig. S8 (a). Let the wave functions at $(k_x, k_y)$, $(k+dk_x, k_y)$, $(k+dk_x, k_y+dk_y)$ and $(k, k_y+dk_y)$ be denoted by $\psi_1, \psi_2, \psi_3$ and $\psi_4$ respectively as shown. The quantity $F_n$ for this $(k_x, k_y)$ is then given by

$$F_n(\boldsymbol{k}) = -i \, \log(\langle \psi_1|\psi_2\rangle \langle \psi_2|\psi_3\rangle \langle \psi_3|\psi_4\rangle \langle \psi_4|\psi_1\rangle), \tag{S16}$$

and the valley Chern number will then be

$$C_n^V = \frac{1}{2\pi} \sum_k F_n(\boldsymbol{k}). \tag{S17}$$

The total valley Chern number is obtained by summing over all the occupied bands, namely, the valence bands, since the Fermi level lies in the band gap. Due to time-reversal symmetry, we find that $C^K = C^{K'}$. We find numerically that the valley Chern number $C^V$ for a particular valley has a finite value for all values of $\Delta$. The value of $C^V$ at $K$ point changes from 2.5 to $-0.5$ close to $\Delta = 0.27$ eV as shown in Fig. S8 (b). This matches with the theoretical estimate of $\sigma_{xy}^V$ in Ref. 18. Our region of interest lies is the range of $\Delta$ from 0.02 eV to 0.05 eV where we have a finite $C^V$ of 2.5 giving $\sigma_{xy}^V = -(e^2/\hbar)C^V = 2.5(e^2/\hbar)$. This suggests that there is a possibility of having edge modes in the system. However, they would not be robust to perturbations since the counter-propagating modes from $K$ and $K'$ valley can hybridize. In the next section, we will explicitly look for edge modes in trilayer graphene within the range of $\Delta$ of interest.

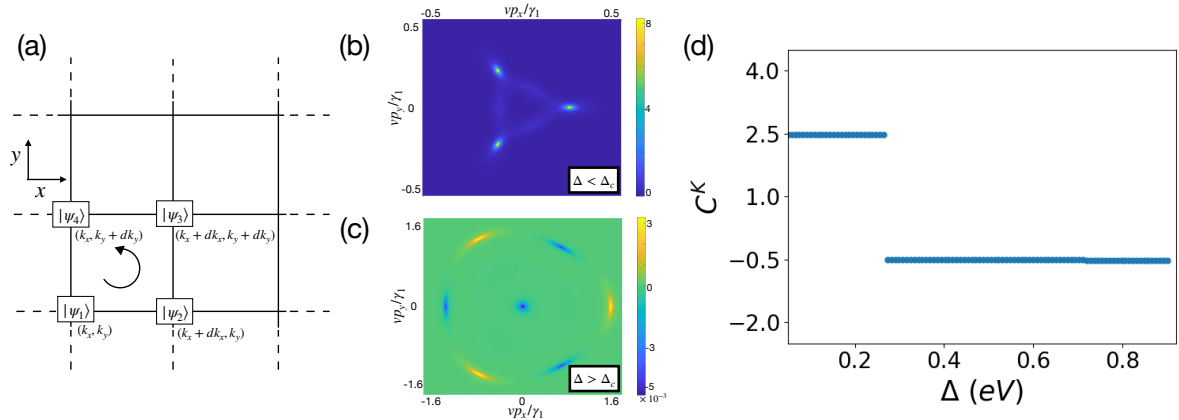

**Fig. S 8:** (a) The discretized mesh grid of the $k$ space used to evaluate the Chern number using Fukui's method. The quantity $F_n(\bm{k})$ as defined in Eq. (S16) is then summed over all these plaquettes in the chosen region of $k$ space. The region is chosen such that all the contribution coming from the Berry curvature lie within this region. As we can see from in Figure (b), the contribution from the $\bm{k}$ lying outside this region is zero. (b) Surface plot of the Berry curvature for the third band (i.e., the valence band) of the system for $\Delta = 0.05$ eV, which is less than the critical field. We see that the contribution to the Berry curvature rapidly goes to zero away from the six off-center Dirac points. (c) Surface plot of the Berry curvature of the valence band for $\Delta = 0.7$ eV. Once again, the maximum contribution comes from the six off-centered Dirac points.

**Presence of edge modes in trilayer graphene at finite $\Delta$:** As presented above, the experimental data suggest linear scaling between the local and non-local resistance. This indicates the presence of edge modes in the system, as will be discussed further in SM Sec. 7. We will first investigate theoretically if trilayer graphene can indeed host edge modes at the values of $\Delta$ of our interest. We implement the method described by T. Morimoto et al. in their work[18]. We consider a semi-infinite system with a zig-zag boundary along the $x$-direction. The edge modes on the left and right edges are given by the regions $y > 0$ and $y < 0$, respectively. We note that the presence of armchair edge dilutes the contribution from edge transport as it enhances valley mixing in the presence of atomic-size scatterers. Since the system is periodic in the $x$-direction, $p_x$ is a good quantum number. However along the $y$-direction, we need to replace $p_y$ by the corresponding operator $\hat{p}_y = -i\hbar\frac{\partial}{\partial y}$. The Hamiltonian can now be divided into parts dependent on and independent of $p_x$, as $\hat{H} = A\hat{p}_y + B(p_x)$. In the basis defined in Eq. (S6), these matrices have the form

$$A = \begin{pmatrix} 0 & 0 & 0 & -iv_0 & 0 & 0 \\ 0 & 0 & -i\sqrt{2}v_4 & 0 & iv_0 & -i\sqrt{2}v_3 \\ 0 & i\sqrt{2}v_4 & 0 & 0 & 0 & iv_0 \\ iv_0 & 0 & 0 & 0 & 0 & 0 \\ 0 & -iv_0 & 0 & 0 & 0 & -i\sqrt{2}v_4 \\ 0 & i\sqrt{2}v_3 & -iv_0 & 0 & i\sqrt{2}v_4 & 0 \end{pmatrix},$$

13

$$B = \begin{pmatrix} -\gamma_2/2 & 0 & 0 & v_0 p_x & 0 & \Delta \\ 0 & 0 & -\sqrt{2}v_4 p_x & 0 & v_0 p_x & \sqrt{2}v_3 p_x \\ 0 & -\sqrt{2}v_4 p_x & \gamma_5/2 + \Delta' & \Delta & \sqrt{2}\gamma_1 & v_0 p_x \\ v_0 p_x & 0 & \Delta & -\gamma_5/2 + \Delta' & 0 & 0 \\ 0 & v_0 p_x & \sqrt{2}\gamma_1 & 0 & \Delta' & -\sqrt{2}v_4 p_x \\ \Delta & \sqrt{2}v_3 p_x & v_0 p_x & 0 & -\sqrt{2}v_4 p_x & \gamma_2/2 \end{pmatrix}.$$

The Schrödinger equation $\hat{H}\psi = \epsilon\psi$ then gives

$$\frac{\partial}{\partial y}\psi = iA^{-1}(\epsilon - B)\psi = C\psi \tag{S18}$$

where $C$ is an eigenvalue of $\partial/\partial y$. For a given $p_x$ with an energy lying within the bulk gap, we look for a wave function of the form $\exp(p_{y,n}y)u_n$, where $p_{y,n}$ and $u_n$ are the complex eigenvalues and eigenvectors of Eq. (S18). For this $6 \times 6$ Hamiltonian, we have six eigenvalues for every pair of values $(p_x, \epsilon)$. Out of these, three eigenvalues have $\text{Re}(p_y) > 0$ and the other three have $\text{Re}(p_y) < 0$. For modes at the right edge, lying in the region $y < 0$, we choose the eigenvalues for which $\text{Re}(p_y) > 0$ so that our wave function decays into the system as we go away from the edge. For modes at the left edge, we take the eigenvectors which have $Re(p_y) < 0$. The six-component general wave function for either region thus looks like

$$\psi = \sum_{n=1}^{3} C_n U_n, \quad \text{where} \quad U_n = \exp(p_{y,n}y)u_n. \tag{S19}$$

From Fig. S10 (a), we can see that the atoms from the $B$ sublattice form the right edge. This puts a condition on the wave function of the atoms on the $A$ sublattice and their components in the wave function are set to zero. Thus we have $|A_1\rangle = 0$, $|A_2\rangle = 0$ and $|A_3\rangle = 0$. This implies that in the re-defined basis, the wave function in Eq. (S19) must satisfy the constraint

$$\begin{pmatrix} U_1^1 & U_2^1 & U_3^1 \\ U_1^5 & U_2^5 & U_3^5 \\ U_1^6 & U_2^6 & U_3^6 \end{pmatrix} \begin{pmatrix} C_1 \\ C_2 \\ C_3 \end{pmatrix} = \begin{pmatrix} 0 \\ 0 \\ 0 \end{pmatrix}, \tag{S20}$$

where $U_n^j$ is the $j^{th}$ component of the $n^{th}$ eigenvector and is a function of $p_x$ and $\epsilon$. For a given $p_x$, we look for values of $\epsilon$ within the band gap such that the determinant of the matrix involving $U_n^j$ is zero. Similarly, we calculate the allowed values of $\epsilon$ for the left edge. In this case, we consider eigenvalues with $Re(p_y) < 0$ with the boundary condition $|B_1\rangle = 0$, $|B_2\rangle = 0$ and $|B_3\rangle = 0$.

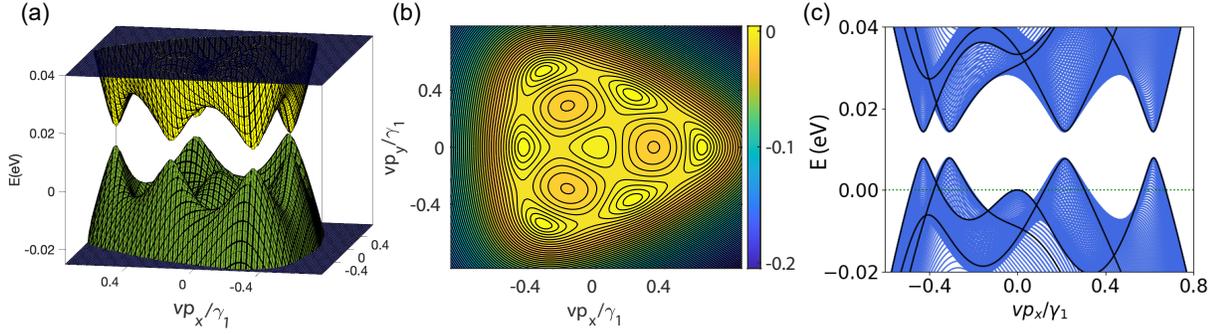

**Fig. S 9:** (a) Three-dimensional plot of the dispersion of TLG for $\Delta = 0.2$ eV close to the Dirac point. (b) Contour plot for the same quantity showing the six off-centred Dirac points around the $K$ point marked with black circles. (c) The bulk dispersion for $\Delta = 0.2$ eV near the Fermi level as a function of $p_x$ along lines passing through the six Dirac points.

The displacement field used in the experimental measurements goes from $D = 0.25$ V/nm to $0.5$ V/nm. The potential difference between the top and the bottom layer is $2\Delta = -(\frac{d_\perp}{\epsilon_{TLG}} \times D)e$, where $d_\perp$ nm is the distance between the top and bottom layers of TLG, $\epsilon_{TLG}$ is the dielectric constant of the material, and $e$ is the electronic charge. Substituting the literature values of $d_\perp = 0.67$ nm and $\epsilon_0 = 4$, we find that the corresponding range of $\Delta$ goes from $20$ meV to $40$ meV. In Figs. S10 (b) and (c), we have plotted the left (red) and right (green) edge modes for $\Delta = 10$ meV and $\Delta = 40$ meV. For values of $\Delta$ less than $20$ meV, we notice that these edge modes become less dispersive and might not effectively contribute to transport. This supports the experimental data where the exponent $\alpha$ deviates from $1$ for smaller values of $D$.

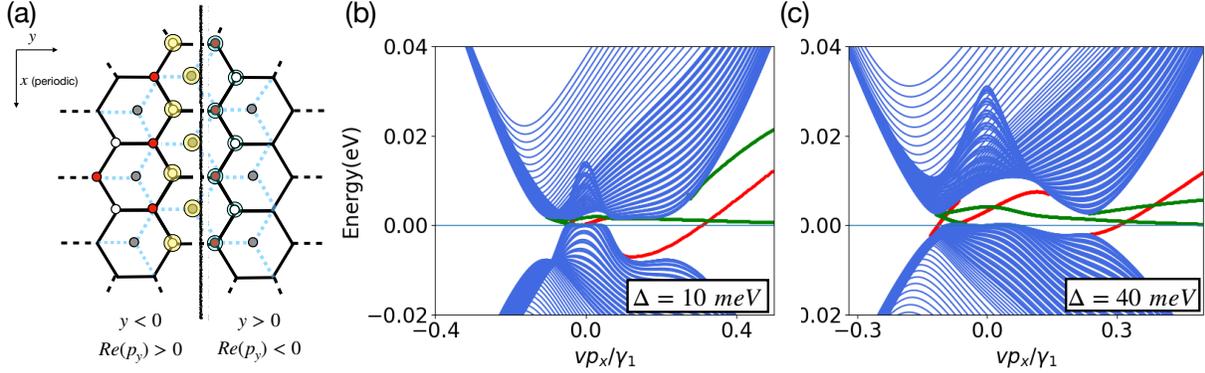

**Fig. S 10:** (a) Zig-zag edge for trilayer graphene. The yellow-coloured atoms $B_1, B_2$ and $B_3$ in the region $y < 0$ form the right edge. The blue circled atoms $A_1, A_2$ and $A_3$ in the region $y > 0$ form the left edge. (b) The dispersion of trilayer graphene, along with the energies of the zig-zag edge modes, are plotted. The green (red) curves correspond to modes on the right (left) edge. This plot is for $\Delta = 10$ meV. (c) The same plot shows edge modes within the band gap for $\Delta = 40$ meV.

**SM section-7: Various origins for non-local signal and their contributions**

In this section, we discuss the few mechanisms that are believed to be possible origins of non-local signals measured in graphene-based devices. (1) Classical ohmic contribution which can be calculated with the well-known formula $R_{NL} = \frac{W R_L}{\pi L}\exp(-\pi L/W)$, [17,21,22] where $L$ is the separation between the current and voltage probe and $W$ is width of the device. (2) The non-local signal originating from the bulk valley current due to the Berry curvature hot spots. This mechanism has been used to explain the findings reported in graphene/hBN superlattice[21] and gapped bilayer graphene[16,17] devices. (3) The edge-mediated non-local charge transport due to topological[18,23,24] or non-topological (charge accumulation at the physical edge of the sample)[25] edge modes. These three mechanisms have also been highlighted in the recent articles [26,27].

    The classical ohmic contribution can be estimated using the well-known equation mentioned above and is ruled out in our experiment as its contribution is orders of magnitude smaller than the observed non-local signal. The second mechanism (bulk valley current) is under debate. Theoretical work reported in Ref. 28 pointed out that non-local resistance peaks cannot be taken as a signature of the anomalous velocities produced by an electric field and a non-vanishing Berry curvature. A crucial point leading to this conclusion was that the use of a semi-classical theory needs to be clarified because, in graphene/hBN superlattice and gapped bilayer graphene experiments, the Fermi level was tuned to lie in an electronic gap, where transport occurs in the tunnelling regime and a semi-classical formalism should not be used.

    Now we discuss the third mechanism (edge-mediated transport), which is believed to be the possible

origin of the observed non-local signal in graphene devices as supported by several theoretical articles[23,24]. However, it should be noted that these theoretical articles do not consider the charge accumulation effect (non-topological edge) as the origin of the non-local signal, as reported for the non-aligned single-layer graphene device[25].

Non-topological (charge accumulation) origin of non-local resistance: The magnitude of the non-local resistance due to the charge accumulation at the physical boundary of the sample can be calculated using the equation [25]

$$R_{NL} = \frac{W R_L}{\pi L} \exp\left(-\frac{\pi L}{\lambda}\right), \tag{S21}$$

with

$$\lambda = W \sqrt{1 + \frac{2\eta\left(1 + \frac{\pi^2}{4}\left(\frac{1+\eta(\mu^2 B^2+1)}{1+\eta}\right)^2\right)}{1 + 2\eta\frac{1+\eta(\mu^2 B^2+1)}{1+\eta}}}. \tag{S22}$$

where $\eta$, $\mu$ and $B$ are the charge accumulation factor, mobility, and magnetic field, respectively.

As shown in Figure 2 (c) of the main manuscript, the estimated value of the non-local resistance due to the charge accumulation factor of $\eta = 0.7$ (this is the typical charge accumulation factor reported in Ref. 25), $R_{ch}$, is one order of magnitude smaller than the observed non-local signal in our device. Further, the linear decay of $R_{NL}$ with $L$ (Fig. S5 (c)) rules out the charge accumulation contribution, which would have scaled exponentially with the length as seen in eqn. S21[25]. The magnitude and its length dependence provide us with enough grounds to firmly believe that the significant contribution to the non-local signal observed in our devices is due to the dispersive edge modes predicted to exist in Bernal stacked trilayer graphene under the application of a sufficient displacement field[18] and is not due to the charge accumulation effect [25].

The contribution from topological edge modes and non-topological edge modes (due to charge accumulation) can co-exist in an edge-mediated transport mechanism. However, in the work of Aharon-Steinberg et al[25], the single-layer graphene sample was not aligned with hBN. Naively, one does not expect the presence of dispersive edge modes in single-layer graphene devices. Some edge modes can exist along the zig-zag edge of single-layer graphene[29]. However, these edge modes are flat (non-dispersive) and do not contribute to the transport. This is probably why charge accumulation was the dominant source of non-local transport reported in Re. 25. As pointed out in Refs. 23, 24, the alignment with hBN (as in Ref. 21) helps these edge modes to gain a dispersion effectively, and then they start contributing to the charge transport. This shortcoming is further discussed in Ref. 27.

17

**SM section-8: Thermal cycle**

To strengthen the robustness and reproducibility of our data shown in the main manuscript, we perform similar non-local measurements in a different thermal cycle of the device. A thermal cycle is meant to expose the device from the low temperature to ambient temperature and atmospheric environment and further cool down the device to low temperature from the room temperature in a different cool-down process. A thermal cycle changes the disorder configurations and impurity concentration in graphene devices and, in principle, can be treated as an entirely new device from the disorder configuration perspective. The results of the local and non-local resistance measurements at $T = 1.8$ K temperature after the second cooldown are summarized in Fig. S 11. Fig. S 11 (a, b) shows the color map of the local resistance ($R_L$) and non-local resistance ($R_{NL}$) as a function of the displacement field $D$, and total density $n$. In Fig. S 11 (c), we plot $\ln R_{NL}$ versus $\ln R_L$ as a function of $n$ for different values of $D$ from $-0.20$ to $-0.50$ V/nm. The data points for these plots are extracted along the horizontal dashed yellow arrows shown in Fig. S 11 (a,b). The inset of Fig. S 11 (c) shows $\ln R_{NL}$ versus $\ln R_L$ as a function of $D$ for the different values of $n$ close to the Dirac point. The data points for the inset plot are extracted along the vertical dashed black arrows shown in Fig. S 11 (a,b). The scaling analysis of Fig. S 11 (c) and its inset show that linear fitting of the plot $\ln R_{NL}$ versus $\ln R_L$ gives a slope equal to one ($\alpha \approx 1$). Fig. S 11 (d) shows the scaling exponent ($\alpha$) as a function of the $D$. Scaling exponent ($\alpha$) remains close to one only above a critical $|D| \gtrsim 0.2$ V/nm, similar to the plot shown in Fig. 3(a) of the main manuscript obtained during the first cooldown of the device. The data in Fig. S 11 shows the robustness and reproducibility of the experimental results in two different cooldown processes.

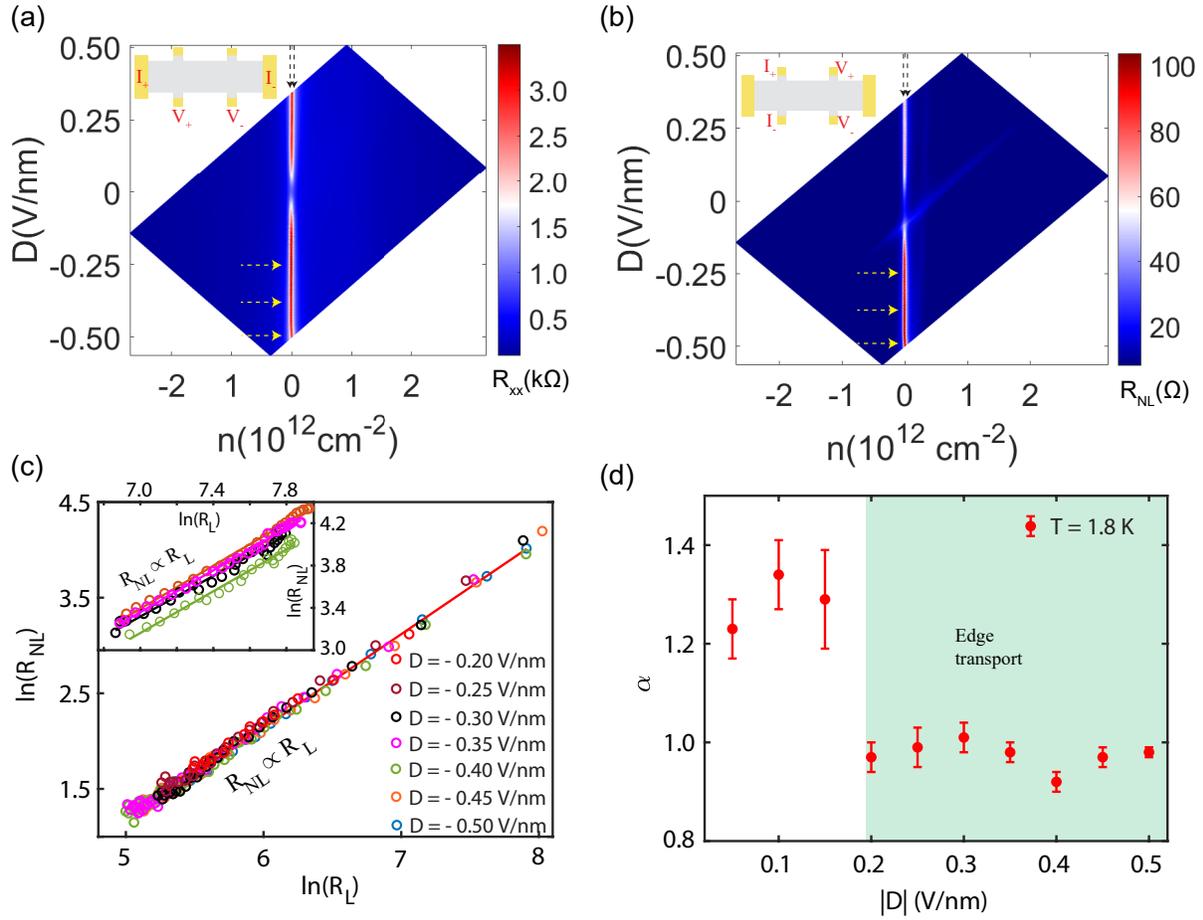

**Fig. S 11: Non-local charge transport in different thermal cycles of the device** Color map of $R_L$ (a) and $R_{NL}$ (b) as a function of total carrier density $n$ and the displacement field $D$ at $T = 1.8$ K. (c) Log-log plots of $R_{NL}$ with $R_L$. Open circles are extracted from Figs.(a, b) for different values of $D$ (along horizontal yellow arrows). The solid lines correspond to the linear fitting of the data points with slope $\sim 1$. Inset: Log-log plots of $R_{NL}$ with $R_L$. Open circles are extracted from Figs.(a, b) for different values of $n$ (along vertical black arrows) near the Dirac point. (d) The scaling exponent $\alpha$ plotted as a function of $D$ at $T = 1.8$ K. For $|D| < 0.2$ V/nm, $\alpha$ deviates significantly from unity.

19



\end{document}